\input epsf.sty
\documentclass[12pt]{article}

\begin{document}        
\pagestyle{empty}
\renewcommand{\thefootnote}{\fnsymbol{footnote}}

\begin{flushright}
{\small
SLAC--PUB--8195\\
July 1999\\}
\end{flushright}

\vspace{.8cm}

\begin{center}
{\bf\large   
Measurement of $A_c$ with charmed mesons at SLD\footnote{Work supported by
Department of Energy contract  DE--AC03--76SF00515 (SLAC).}}

\vspace{1cm}
{\bf The SLD Collaboration $^{**}$
} \\
\vskip 0.2cm
{\it
Stanford Linear Accelerator Center, Stanford University,
Stanford, CA  94309
}
\medskip
\end{center}
 
\vfill

\begin{center}
{\bf\large   
Abstract }
\end{center}
We present a direct measurement of the parity-violation parameter $A_c$.
The measurement is based on 550k $Z^0$ decays 
collected by the SLD detector.
The mean electron-beam polarization is $|P_e| = 73 \%$.
The tagging of $c$-quark events was performed using two methods:
The exclusive reconstruction of 
$D^{\ast+}$, $D^+$, and $D^0$ mesons, and the inclusive $P_T$ 
spectrum of soft-pions ($\pi_s$) in the decay of
$D^{\ast+}\rightarrow D^0 \pi_s^+$.
The results of these two methods are combined to give
$$A_c = 0.688 \pm 0.035(stat.) \pm 0.025(sys.) \, \mbox{ (preliminary).}$$

\vfill

\begin{center} 
{\it Contributed to the International Europhysics Conference 
on High Energy Physics, July 15-21 1999, Tampere, Finland; Ref 6-474,
and to
the XIXth International Symposium on Lepton 
Photon Interactions, August 9-14 1999, Stanford, USA. }

\end{center}

\newpage


 
 
%
\pagestyle{plain}
\section{ Introduction }

In the Standard Model, the electroweak interaction of $Z^0$ to
fermions 
has both vector ($v$) and axial-vector ($a$) couplings. 
Measurement of fermion asymmetries at the $Z^0$ resonance probe a combination 
of these couplings given by 
$A_f = 2v_f a_f/(v_f^2 + a_f^2)$.
The parameters $A_f$ express the extent of parity violation at the $Zff$ vertex
and provide sensitive tests of the Standard Model.

At the Born-level, the differential cross section 
for the reaction $e^+e^- \rightarrow 
Z^0 \rightarrow f\bar{f}$ is 
$$ \sigma_f(z) \equiv d\sigma_f/dz 
\propto (1-A_eP_e)(1+z^2) + 2A_f(A_e - P_e) z, $$
where $P_e$ is the longitudinal polarization of the electron beam 
and $z = \cos\theta$ is the 
direction of the outgoing fermion relative to the incident electron.
At the SLAC Linear Collider (SLC), the ability to manipulate the 
longitudinal polarization of the electron beam allows the isolation
of the parameter $A_f$ independently 
through formation of the left-right forward-backward
double asymmetry:
$ \tilde{A}_{FB}^f(z) = |P_e| A_f 2z/(1+z^2).$

In this note, we present the direct measurement of the
parity-violation parameter for $c$-quarks, $A_c$.
The tagging of $c$-quarks was performed using
exclusively reconstructed 
$D^{\ast+}$, $D^+$, and $D^0$ mesons, as well as an 
inclusive sample of $D^{\ast+}\rightarrow D^0 \pi_s$ decays
identified by the soft-pion ($\pi_s$).

\section{ Apparatus and event selection }
The measurement described here is based on 
550k $Z^0$ decays recorded in 1993-98
with the SLC Large Detector (SLD) at the SLC $e^+e^-$ collider. 
A general description of the SLD can be found elsewhere \cite{sld}.
For charged-particle tracking devices, 
we use the central drift chamber (CDC)\cite{cdc} and 
a pixel-based silicon vertex detector (VXD)\cite{vxd2,vxd3}.
The Liquid Argon Calorimeter (LAC)\cite{lac} measures the energy of 
charged and neutral particles and is also used for electron 
identification. Muon tracking is provided by the Warm Iron Calorimeter
(WIC)\cite{wic}.
The \v Cerenkov Ring Imaging Detector (CRID)\cite{crid} provides
particle identification. 
The SLC was operated with a polarized electron beam and an unpolarized 
positron beam \cite{SLC}.
The average polarization magnitude measured for
the 1993-98 data sample is $|P_e| = 73 \%$.

Hadronic events are selected 
by requiring at least 5 charged tracks, 
a total visible energy of at least 20 GeV/c, and
a thrust axis calculated from charged tracks satisfying
$|\cos\theta_{thrust}| < 0.87 $.
As $Z^0 \rightarrow b\bar{b}$ events are also a copious source of
$D$ mesons, they represent a potential background.
We reject these events using the invariant mass of
charged tracks associated with 
reconstructed secondary decay
vertices\cite{masstag1, masstag2}.
Specifically, we require that the reconstructed secondary vertices have 
mass less than 2.0 GeV/c$^2$. 
Our simulations show that 
this cut rejects 57\% of 
$b\bar{b}$ events with 99\% of the remainder being $c\bar{c}$ events.

\section{ $A_c$ measurement with exclusive charmed-meson
reconstruction} 
%
$D^{\ast+}$ mesons are identified 
via their decay $D^{\ast+} \rightarrow \pi_s^+ D^0 $ 
followed by: 
\[\begin{array}{ll} 
 D^0 \rightarrow K^- \pi^+  & ``K\pi'',\\       
 D^0 \rightarrow K^- \pi^+ \pi^0 & ``Satellite'',\\
 D^0 \rightarrow K^- \pi^+ \pi^- \pi^+ & ``K\pi\pi\pi'', \mbox{ or} \\
 D^0 \rightarrow K^- l^+ \nu_l \hskip 0.5cm \mbox{($l$=e or $\mu$)} & ``Semileptonic''.
\end{array}\]
To form the $D^{\ast+}$ candidate, we use all tracks which 
have VXD hits in one event
hemisphere.
For the $D^0$ candidate, a number of tracks corresponding to the
charged multiplicity in each $D^0$ decay mode are combined 
with assuming one of them to be a kaon and the others are pions.
Only candidates with the correct charge combinations are selected.
A vertex fit is performed on the tracks in a candidate, and 
we require that the combination has a $\chi^2$ probability of
all tracks coming from the same vertex be greater than 1\%.
Then if the calculated invariant mass lies within 
the mass ranges of
1.765 GeV/c$^2$ $<$ m$_{D^0}$ $<$ 1.965 GeV/c$^2$ ($K\pi$),
1.500 GeV/c$^2$ $<$ m$_{D^0}$ $<$ 1.600 GeV/c$^2$ ($Satellite$),
1.795 GeV/c$^2$ $<$ m$_{D^0}$ $<$ 1.935 GeV/c$^2$
($K\pi\pi\pi$), and
1.100 GeV/c$^2$ $<$ m$_{D^0}$ $<$ 1.800 GeV/c$^2$ ($Semileptonic$),
%
combine with a soft-pion candidate track which has a charge
opposite to the kaon candidate, to form the
$D^{\ast+}$ candidate.

\begin{figure}
\epsfysize14cm
\hskip0.5in
\epsfbox{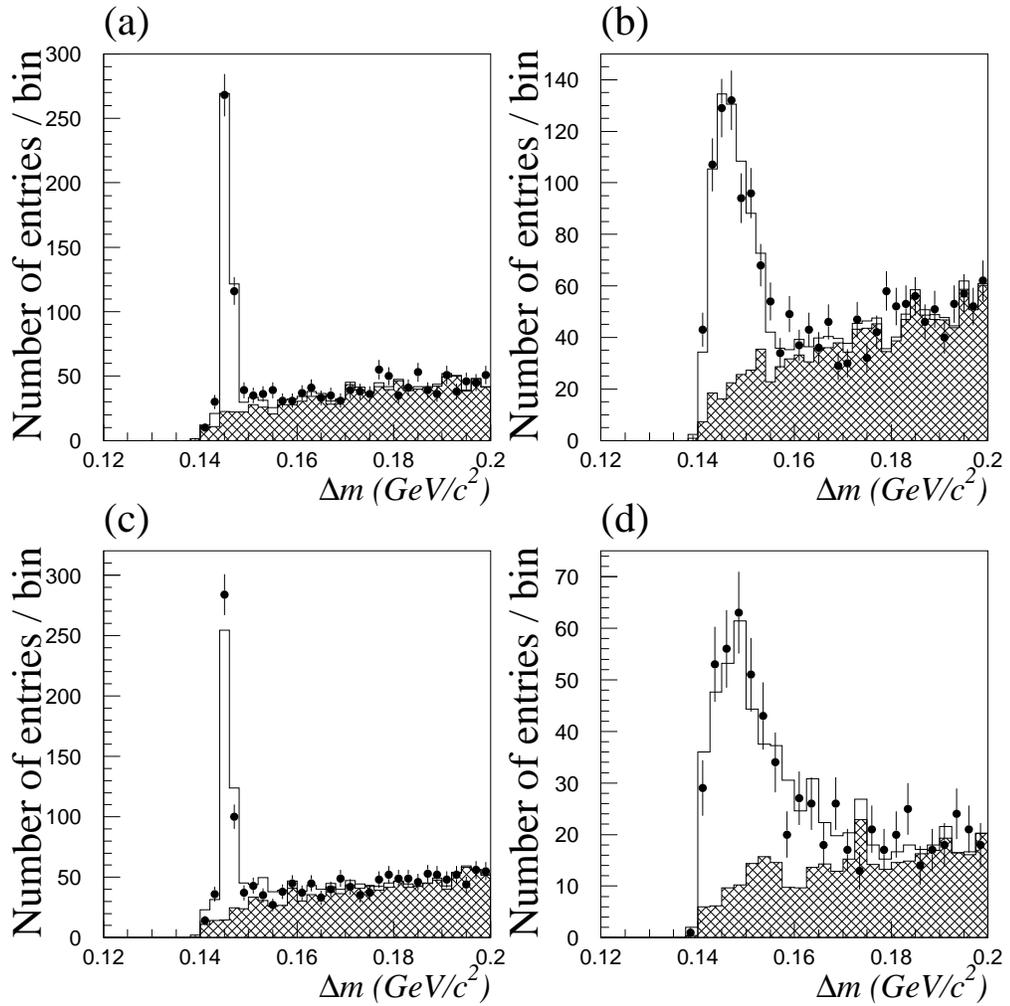}
\vskip -0.5cm
\caption{The mass-difference distributions for the decay of
(a)$D^{\ast+}\rightarrow D^0 \pi_s$, $D^0 \rightarrow K\pi$, 
(b)$D^0 \rightarrow K\pi\pi^0$, 
(c)$D^0 \rightarrow K\pi\pi\pi$, and  
(d)$D^0 \rightarrow K^- l^+ \nu_l$ ($l=$e or $\mu$).
The solid circles indicate the exp. data, and histograms are
MC of signal (open) and RCBG (hatched).}
\label{fig:dmass}
\end{figure}

To reconstruct the $D^{\ast}$, we use two sets of selection criteria.
One is based on event kinematics and the other on event topology. 
We select the combinations which satisfy either of the two.
In the former one,
we require the candidate to have $x_{D^{\ast}} > 0.4$
($K\pi$, $Satellite$, and $Semileptonic$) or 0.6 ($K\pi\pi\pi$),
where $x_{D^{\ast}}\equiv 2E_{D^{\ast}}/E_{CM}$, and
$|\cos\theta^{\ast}| < 0.9$ ($K\pi$, $Satellite$, and $Semileptonic$)
or 0.8 ($K\pi\pi\pi$),
where $\theta^{\ast}$ is the opening
angle between the direction of the $D^0$ in the lab. frame and 
the kaon in the $D^0$ rest frame.
We also require the soft-pion candidate to have momentum
greater than 1 GeV/c.
%
In the selection based on the event topologies, 
we require the reconstructed  $D^0$ vertices to have 
decay length $L/\sigma_L > 2.5$, and 
the $xy$ impact parameter of the $D^0$ momentum
vector to the IP to be less than 20 $\mu$m ($K\pi$ and $K\pi\pi\pi$) or
30 $\mu$m ($Satellite$ and $Semileptonic$).
Finally, a cut of $x_{D^{\ast}}>0.3$ ($K\pi$,
$Satellite$, and $Semileptonic$) 
or 0.4 ($K\pi\pi\pi$) is applied.
%
%
Then the mass difference $\Delta M = M_{D^{\ast}} - M_{D^0}$ is
formed. 
We regard the combination as a signal when 
$\Delta M <$ 0.148 GeV/c$^2$ ($K\pi$ and $K\pi\pi\pi$), $<$ 0.155 GeV/c$^2$
($Satellite$), or $<$ 0.16 GeV/c$^2$ ($Semileptonic$).
The side-band region is defined as 0.16 $< \Delta M <$ 0.20 GeV/c$^2$
 or 0.17 $< \Delta M <$ 0.20 GeV/c$^2$ ($Semileptonic$),
to estimate random combinatoric background (RCBG) 
contamination in the signal region.
The mass difference spectra for the four reconstructed $D^{\ast +}$
decay modes are shown in Fig.~\ref{fig:dmass}. 
%

%
The $D^+$ and $D^0$ mesons are identified via the decay of:
\[\begin{array}{l}
D^+ \rightarrow K^- \pi^+ \pi^+ \mbox{ and} \\        
D^0 \rightarrow K^- \pi^+.
\end{array}\]       
For the $D^+$ reconstruction, we combine two same-sign pion candidates 
with an opposite-sign kaon candidate.
We apply the cuts  $x_{D^+} > 0.4$, and 
$\cos\theta^{\ast} > -0.8$.
We require that the $\chi^2$ probability of the good vertex fit be 
$>$ 1$\%$, and apply the $D^+$ decay length cut of $L/\sigma_L > $3.0.
The collinearity angle between the $D^+$ momentum vector and
the vertex flight direction is required 
to be less than 5 mrad in $xy$ and less than 20 mrad in $rz$.

\begin{figure}
\epsfysize7.7cm
\hskip0.5in
\epsfbox{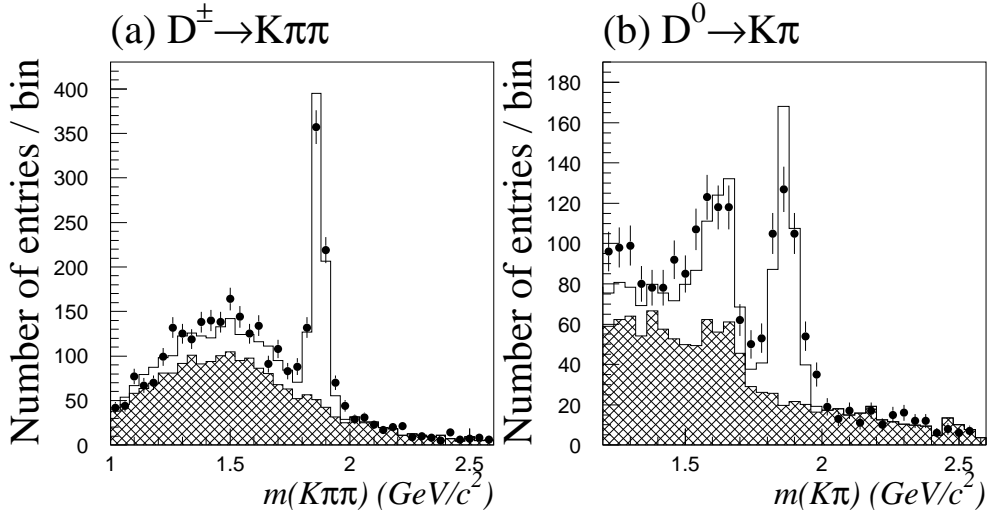}
\vskip -0.5cm
\caption{The mass distributions for (a)$D^+$ and (b)$D^0$ mesons.
The solid circles indicate the exp. data, and histograms are
the MC of signal (open) and RCBG (hatched).}
\label{fig:d+0mass}
\end{figure}

To form the $D^0$ vertices, kaon-candidate tracks identified 
with the CRID are combined with another opposite-sign pion candidate.
We require $x_{D^0} > 0.4$,
$\chi^2$ probability $>$ 1$\%$, and $L/\sigma_L > $3.0.
Finally, 
we require the $xy$ impact parameter of the $D^0$ momentum
vector to the IP to be less than 20 $\mu$m.
The $D^+$ and $D^0$ candidates in the ranges of 
1.800 $< m(K^-\pi^+\pi^+) <$ 1.940 GeV/c$^2$ and 
1.765 $< m(K^-\pi^+) <$ 1.965 GeV/c$^2$ are regarded as signals, respectively.
The sideband regions are defined as 
1.640 $< m(K^-\pi^+\pi^+) <$ 1.740 GeV/c$^2$ and 
2.000 $< m(K^-\pi^+\pi^+) <$ 2.100 GeV/c$^2$ for $D^+$, and
2.100 $< m(K^-\pi^+\pi^+) <$ 2.500 GeV/c$^2$ for $D^0$.
In the Fig.~\ref{fig:d+0mass},  the invariant mass spectra for
the $D^+$ and $D^0$ are plotted.

From 1993-98 SLD data, we select 3967 candidates which 
consist of 2829  $c \rightarrow D$ signal, 
281 $b \rightarrow D$, and 857 RCBG.

\begin{figure}
\epsfysize10cm
\hskip1.in
\epsfbox{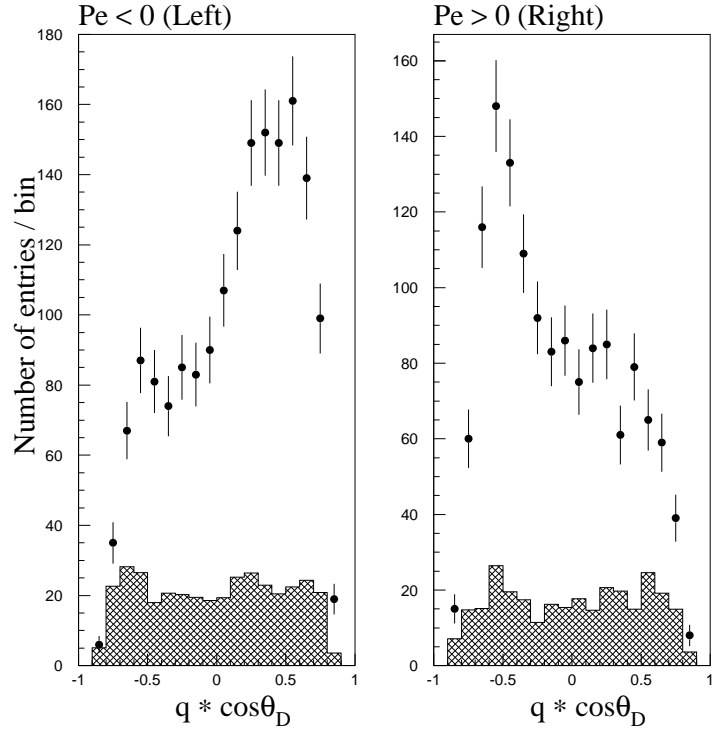}
\vskip -0.5cm
\caption{The distributions of the $q \cdot \cos\theta_D$ for
the selected $D$ meson sample for (a) left- and (b) right-
handed electron beams.
The solid circles are exp. data, and hatched histograms are RCBG
estimated from side-band regions.}
\label{fig:qcos_ex}
\end{figure}

The charge of the primary $c$-quark is determined 
by the sign of the $D^{(\ast)+}$, or  $K^-$ (in the $D^0$ case).
The direction of the primary quark is estimated from 
the direction of the reconstructed $D$ meson.
Fig.~\ref{fig:qcos_ex} shows $y \equiv q \cos\theta_D$ distributions 
for the selected $D$ meson sample separately for left- and right-handed
electron beams. 

To extract $A_c$, we use an unbinned maximum likelihood fit based on
the Born-level cross section for fermion production in $Z^0$-boson
decay. 
The likelihood function used in this analysis is
\begin{eqnarray}
\ln{\cal L}= \sum^{n}_{i=1} 
& \ln & \{
P_c^j(x_D^i) \cdot [(1-P_eA_e)(1+y_i^2)
+2(A_e-P_e)y_i \cdot A_c^D 
\cdot (1-\Delta_{QCD}^c(y_i))
] \nonumber \\
 & + &  
P_b^j(x_D^i) \cdot [(1-P_eA_e)(1+y_i^2)
+2(A_e-P_e)y_i\cdot A_b^D 
\cdot (1-\Delta_{QCD}^b(y_i))
] \nonumber \\
 & + &  
P_{RCBG}^j(x_D^i) \cdot [(1+y_i^2)
+2A_{RCBG}y_i]
\}
\nonumber 
\end{eqnarray}
where 
$n$ is the total number of candidates,
and $A_c^D$ and $A_b^D$ are the asymmetries from $D^{\ast}$,
$D^+$, and $D^0$ mesons in $c\bar{c}$ and $b\bar{b}$ events, 
respectively. 
The index $j$ indicates each of the six charm decay modes.
For the asymmetry for the RCBG, 
we take $A_{RCBG} = 0$ for the central value.  
For $A_e$, we have taken 
$A_e$ = 0.1512 $\pm$ 0.0043 from  the SLD measurement\cite{SLDAe}.
$P_c^j$, $P_b^j$,  and $P_{RCBG}^j$ are the probabilities 
that a candidate from the $j$th decay mode 
is a signal from $c\bar{c}$, $b\bar{b}$,
or RCBG.
$\Delta_{QCD}^f(y)$ is the $O(\alpha_s)$ QCD correction to the
asymmetry. 
Only $A_c^D$ is treated as a free parameter, and
$A_b^D$ is taken to be fixed.

Performing the maximum likelihood fit to the data sample, we obtain
$$A_c = 0.690 \pm 0.042,$$
where the error is statistical only.
%
\section{ Inclusive soft-pion analysis}

In this analysis, $c$-quarks are identified using 
an inclusive soft-pion.
Since the decay $D^{\ast+}\rightarrow D^0 \pi_s$ has a small
Q value of $m_{D^\ast} - m_{D^0} - m_{\pi}$ = 6 MeV$/c^2$,
the maximum transverse momentum of $\pi_s$ with respect to the 
$D^{\ast}$ flight direction is only 40 MeV.

To determine the $D^{\ast}$ direction, charged tracks and neutral
clusters are clustered into jets, using an invariant-mass
algorithm 
where particles are merged together in
an iterative way if their invariant mass is less than 4.6 GeV/c$^2$.
The jets which satisfy 1) at least 3 charged tracks, 2) at least
one track has a momentum of P $>$ 5 GeV/c, 3) a net charge of the 
jet $\le$ 2e,
4) sum of the normalized impact parameters of the largest and
  second largest tracks $>$ 2.5 $\sigma$, and
5) there is at least one opposite-charged-tracks pair which has the 
   $\chi^2$ probability of two tracks coming from the same
   vertex of greater than 1\%,
are selected as $D^{\ast}$ jets.
The transverse momentum $P_T$ to the $D^{\ast}$ jet axis 
for the soft-pion tracks, which have a momentum of $1 < P < 3$ GeV/c and 
an impact parameter of $< 2\sigma$, are calculated. 
\begin{figure}
\epsfysize7.7cm
\hskip1.6in
\epsfbox{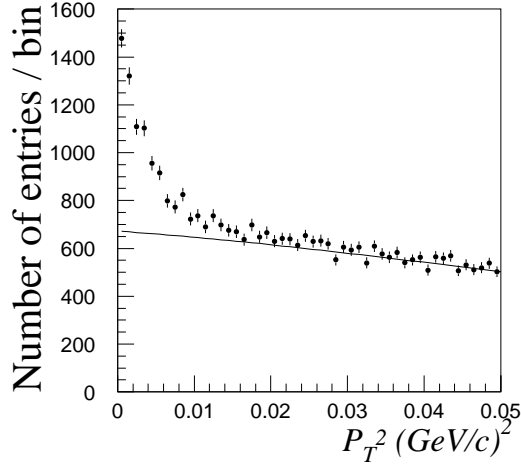}
\vskip -0.5cm
\caption{The $P_T^2$ distribution for soft-pion tracks.
The solid circles indicate the exp. data. Background shape is
obtained by the function described in the text.}
\label{fig:spi_pt}
\end{figure} 
Fig.~\ref{fig:spi_pt} shows the $P_T^2$ distribution for the soft-pion
tracks. 
The background shape is determined by the function of 
$F_{BG}(P_T^2) = a / (1 + bP_T^2 + c(P_T^2)^2)$.
The region of $P_T^2 < 0.01$ (GeV/c)$^2$ is regarded as a
signal region, where a signal-to-background ratio of 1:2 is observed.
From 1993-98 data, 12992 soft-pion tracks are selected 
which consist of 3791 $c \rightarrow D^{\ast+}$, 
500 $b \rightarrow D^{\ast+}$, and 8701 background tracks.

To extract $A_c$, we also use an unbinned maximum likelihood fit
discussed in the exclusive analysis.
We obtain 
$$A_c = 0.683 \pm 0.052(stat.)$$
for the inclusive-soft-pion analysis.

\section{ Results }
Possible systematic errors have been estimated 
and are summarized in Table~\ref{tab:systematic}.
\begin{table}
\begin{center}
\newcommand{\lw}[1]{\smash{\lower2.0ex\hbox{#1}}}
\begin{tabular}{l|c c}
\hline \hline
Systematic error source & Exclusive & Inclusive\\
\hline 
$f_{RCBG}$                     & 0.0111 & 0.0244 \\
Choice of BG shape             & --     & 0.0385 \\
$A_{RCBG}$                     & 0.0028 & 0.0093 \\
RCBG x distribution            & 0.0112 & 0.0018 \\
RCBG acceptance                & 0.0087 & 0.0122 \\
$f_{b \rightarrow D}/(f_{b \rightarrow D}+f_{c \rightarrow D})$ 
                               & 0.0011 & 0.0018 \\
$A_{b \rightarrow D}$ $(A_b)$  & 0.0033 & 0.0040 \\
$A_{b \rightarrow D}$ (Mixing) & 0.0099 & 0.0122 \\
$<x_c>$                        & 0.0003 & 0.0020 \\
$<x_b>$                        & 0.0003 & 0.0010 \\
d($x_D$) shape                 & 0.0040 & 0.0003 \\
$A_e$                          & 0.0004 & 0.0011 \\
Polarization                   & 0.0051 & 0.0047 \\
CRID Kaon ID                   & 0.0022 & --     \\
$\alpha_s$                     & 0.0004 & 0.0005 \\
Correction to QCD corr.        & 0.0018 & 0.0024 \\
2nd Order QCD corr.            & 0.0006 & 0.0008 \\
Gluon splitting                & 0.0001 & 0.0004 \\
\hline
Total   & 0.0222 & 0.0502 \\
\hline \hline
\end{tabular}
\end{center}
\caption{Contributions to the estimated systematic error.}
\label{tab:systematic}
\end{table}
We get the preliminary results of the parity-violation parameter
$A_c$ using two kinds of $c$-quark tagging methods:
$$A_c = 0.690 \pm 0.042(stat.) \pm 0.022(sys.) \mbox{ and}$$
$$A_c = 0.683 \pm 0.052(stat.) \pm 0.050(sys.),$$
from exclusive charmed-meson reconstruction and inclusive soft-pion
analysis, respectively.
To combine them, we avoided double counting signal events from both samples.
1182 events are counted as overlap events in the analyses.
This increased the statistical error for soft-pion analysis to 
0.061.

The combined result from the two analyses is
$$A_c = 0.688 \pm 0.035(stat.) \pm 0.025(sys.) \,\mbox{ (preliminary).}$$

\section*{Acknowledgments}
We thank the personnel of the SLAC accelerator department and the
technical
staffs of our collaborating institutions for their outstanding efforts
on our behalf.

\vskip .5truecm

\vbox{\footnotesize\renewcommand{\baselinestretch}{1}\noindent
$^*$Work supported by Department of Energy
  contracts:
  DE-FG02-91ER40676 (BU),
  DE-FG03-91ER40618 (UCSB),
  DE-FG03-92ER40689 (UCSC),
  DE-FG03-93ER40788 (CSU),
  DE-FG02-91ER40672 (Colorado),
  DE-FG02-91ER40677 (Illinois),
  DE-AC03-76SF00098 (LBL),
  DE-FG02-92ER40715 (Massachusetts),
  DE-FC02-94ER40818 (MIT),
  DE-FG03-96ER40969 (Oregon),
  DE-AC03-76SF00515 (SLAC),
  DE-FG05-91ER40627 (Tennessee),
  DE-FG02-95ER40896 (Wisconsin),
  DE-FG02-92ER40704 (Yale);
  National Science Foundation grants:
  PHY-91-13428 (UCSC),
  PHY-89-21320 (Columbia),
  PHY-92-04239 (Cincinnati),
  PHY-95-10439 (Rutgers),
  PHY-88-19316 (Vanderbilt),
  PHY-92-03212 (Washington);
  The UK Particle Physics and Astronomy Research Council
  (Brunel, Oxford and RAL);
  The Istituto Nazionale di Fisica Nucleare of Italy
  (Bologna, Ferrara, Frascati, Pisa, Padova, Perugia);
  The Japan-US Cooperative Research Project on High Energy Physics
  (Nagoya, Tohoku);
  The Korea Research Foundation (Soongsil, 1997).}



\section*{$^{**}$List of Authors} 


%
%
%
\begin{center}
\def\iADEL{$^{(1)}$}
\def\iAOMORI{$^{(2)}$}
\def\iBOLO{$^{(3)}$}
\def\iBRI{$^{(4)}$}
\def\iBRUN{$^{(5)}$}
\def\iBU{$^{(6)}$}
\def\iCINC{$^{(7)}$}
\def\iCOLO{$^{(8)}$}
\def\iCOLU{$^{(9)}$}
\def\iCSU{$^{(10)}$}
\def\iFERR{$^{(11)}$}
\def\iFRAS{$^{(12)}$}
\def\iILLI{$^{(13)}$}
\def\iJHU{$^{(14)}$}
\def\iLBL{$^{(15)}$}
\def\iLTU{$^{(16)}$}
\def\iMASS{$^{(17)}$}
\def\iMISSI{$^{(18)}$}
\def\iMIT{$^{(19)}$}
\def\iMOSCOW{$^{(20)}$}
\def\iNAGO{$^{(21)}$}
\def\iOREG{$^{(22)}$}
\def\iOXF{$^{(23)}$}
\def\iPADO{$^{(24)}$}
\def\iPERU{$^{(25)}$}
\def\iPISA{$^{(26)}$}
\def\iRAL{$^{(27)}$}
\def\iRUTG{$^{(28)}$}
\def\iSLAC{$^{(29)}$}
\def\iSOGA{$^{(30)}$}
\def\iSOONG{$^{(31)}$}
\def\iTENN{$^{(32)}$}
\def\iTOHO{$^{(33)}$}
\def\iUCSB{$^{(34)}$}
\def\iUCSC{$^{(35)}$}
\def\iUVIC{$^{(36)}$}
\def\iVAND{$^{(37)}$}
\def\iWASH{$^{(38)}$}
\def\iWISC{$^{(39)}$}
\def\iYALE{$^{(40)}$}

  \baselineskip=.75\baselineskip  
\mbox{Kenji  Abe\unskip,\iNAGO}
\mbox{Koya Abe\unskip,\iTOHO}
\mbox{T. Abe\unskip,\iSLAC}
\mbox{I. Adam\unskip,\iSLAC}
\mbox{T.  Akagi\unskip,\iSLAC}
\mbox{H. Akimoto\unskip,\iSLAC}
\mbox{N.J. Allen\unskip,\iBRUN}
\mbox{W.W. Ash\unskip,\iSLAC}
\mbox{D. Aston\unskip,\iSLAC}
\mbox{K.G. Baird\unskip,\iMASS}
\mbox{C. Baltay\unskip,\iYALE}
\mbox{H.R. Band\unskip,\iWISC}
\mbox{M.B. Barakat\unskip,\iLTU}
\mbox{O. Bardon\unskip,\iMIT}
\mbox{T.L. Barklow\unskip,\iSLAC}
\mbox{G.L. Bashindzhagyan\unskip,\iMOSCOW}
\mbox{J.M. Bauer\unskip,\iMISSI}
\mbox{G. Bellodi\unskip,\iOXF}
\mbox{A.C. Benvenuti\unskip,\iBOLO}
\mbox{G.M. Bilei\unskip,\iPERU}
\mbox{D. Bisello\unskip,\iPADO}
\mbox{G. Blaylock\unskip,\iMASS}
\mbox{J.R. Bogart\unskip,\iSLAC}
\mbox{G.R. Bower\unskip,\iSLAC}
\mbox{J.E. Brau\unskip,\iOREG}
\mbox{M. Breidenbach\unskip,\iSLAC}
\mbox{W.M. Bugg\unskip,\iTENN}
\mbox{D. Burke\unskip,\iSLAC}
\mbox{T.H. Burnett\unskip,\iWASH}
\mbox{P.N. Burrows\unskip,\iOXF}
\mbox{R.M. Byrne\unskip,\iMIT}
\mbox{A. Calcaterra\unskip,\iFRAS}
\mbox{D. Calloway\unskip,\iSLAC}
\mbox{B. Camanzi\unskip,\iFERR}
\mbox{M. Carpinelli\unskip,\iPISA}
\mbox{R. Cassell\unskip,\iSLAC}
\mbox{R. Castaldi\unskip,\iPISA}
\mbox{A. Castro\unskip,\iPADO}
\mbox{M. Cavalli-Sforza\unskip,\iUCSC}
\mbox{A. Chou\unskip,\iSLAC}
\mbox{E. Church\unskip,\iWASH}
\mbox{H.O. Cohn\unskip,\iTENN}
\mbox{J.A. Coller\unskip,\iBU}
\mbox{M.R. Convery\unskip,\iSLAC}
\mbox{V. Cook\unskip,\iWASH}
\mbox{R.F. Cowan\unskip,\iMIT}
\mbox{D.G. Coyne\unskip,\iUCSC}
\mbox{G. Crawford\unskip,\iSLAC}
\mbox{C.J.S. Damerell\unskip,\iRAL}
\mbox{M.N. Danielson\unskip,\iCOLO}
\mbox{M. Daoudi\unskip,\iSLAC}
\mbox{N. de Groot\unskip,\iBRI}
\mbox{R. Dell'Orso\unskip,\iPERU}
\mbox{P.J. Dervan\unskip,\iBRUN}
\mbox{R. de Sangro\unskip,\iFRAS}
\mbox{M. Dima\unskip,\iCSU}
\mbox{D.N. Dong\unskip,\iMIT}
\mbox{M. Doser\unskip,\iSLAC}
\mbox{R. Dubois\unskip,\iSLAC}
\mbox{B.I. Eisenstein\unskip,\iILLI}
\mbox{I.Erofeeva\unskip,\iMOSCOW}
\mbox{V. Eschenburg\unskip,\iMISSI}
\mbox{E. Etzion\unskip,\iWISC}
\mbox{S. Fahey\unskip,\iCOLO}
\mbox{D. Falciai\unskip,\iFRAS}
\mbox{C. Fan\unskip,\iCOLO}
\mbox{J.P. Fernandez\unskip,\iUCSC}
\mbox{M.J. Fero\unskip,\iMIT}
\mbox{K. Flood\unskip,\iMASS}
\mbox{R. Frey\unskip,\iOREG}
\mbox{J. Gifford\unskip,\iUVIC}
\mbox{T. Gillman\unskip,\iRAL}
\mbox{G. Gladding\unskip,\iILLI}
\mbox{S. Gonzalez\unskip,\iMIT}
\mbox{E.R. Goodman\unskip,\iCOLO}
\mbox{E.L. Hart\unskip,\iTENN}
\mbox{J.L. Harton\unskip,\iCSU}
\mbox{K. Hasuko\unskip,\iTOHO}
\mbox{S.J. Hedges\unskip,\iBU}
\mbox{S.S. Hertzbach\unskip,\iMASS}
\mbox{M.D. Hildreth\unskip,\iSLAC}
\mbox{J. Huber\unskip,\iOREG}
\mbox{M.E. Huffer\unskip,\iSLAC}
\mbox{E.W. Hughes\unskip,\iSLAC}
\mbox{X. Huynh\unskip,\iSLAC}
\mbox{H. Hwang\unskip,\iOREG}
\mbox{M. Iwasaki\unskip,\iOREG}
\mbox{D.J. Jackson\unskip,\iRAL}
\mbox{P. Jacques\unskip,\iRUTG}
\mbox{J.A. Jaros\unskip,\iSLAC}
\mbox{Z.Y. Jiang\unskip,\iSLAC}
\mbox{A.S. Johnson\unskip,\iSLAC}
\mbox{J.R. Johnson\unskip,\iWISC}
\mbox{R.A. Johnson\unskip,\iCINC}
\mbox{T. Junk\unskip,\iSLAC}
\mbox{R. Kajikawa\unskip,\iNAGO}
\mbox{M. Kalelkar\unskip,\iRUTG}
\mbox{Y. Kamyshkov\unskip,\iTENN}
\mbox{H.J. Kang\unskip,\iRUTG}
\mbox{I. Karliner\unskip,\iILLI}
\mbox{H. Kawahara\unskip,\iSLAC}
\mbox{Y.D. Kim\unskip,\iSOGA}
\mbox{M.E. King\unskip,\iSLAC}
\mbox{R. King\unskip,\iSLAC}
\mbox{R.R. Kofler\unskip,\iMASS}
\mbox{N.M. Krishna\unskip,\iCOLO}
\mbox{R.S. Kroeger\unskip,\iMISSI}
\mbox{M. Langston\unskip,\iOREG}
\mbox{A. Lath\unskip,\iMIT}
\mbox{D.W.G. Leith\unskip,\iSLAC}
\mbox{V. Lia\unskip,\iMIT}
\mbox{C.Lin\unskip,\iMASS}
\mbox{M.X. Liu\unskip,\iYALE}
\mbox{X. Liu\unskip,\iUCSC}
\mbox{M. Loreti\unskip,\iPADO}
\mbox{A. Lu\unskip,\iUCSB}
\mbox{H.L. Lynch\unskip,\iSLAC}
\mbox{J. Ma\unskip,\iWASH}
\mbox{M. Mahjouri\unskip,\iMIT}
\mbox{G. Mancinelli\unskip,\iRUTG}
\mbox{S. Manly\unskip,\iYALE}
\mbox{G. Mantovani\unskip,\iPERU}
\mbox{T.W. Markiewicz\unskip,\iSLAC}
\mbox{T. Maruyama\unskip,\iSLAC}
\mbox{H. Masuda\unskip,\iSLAC}
\mbox{E. Mazzucato\unskip,\iFERR}
\mbox{A.K. McKemey\unskip,\iBRUN}
\mbox{B.T. Meadows\unskip,\iCINC}
\mbox{G. Menegatti\unskip,\iFERR}
\mbox{R. Messner\unskip,\iSLAC}
\mbox{P.M. Mockett\unskip,\iWASH}
\mbox{K.C. Moffeit\unskip,\iSLAC}
\mbox{T.B. Moore\unskip,\iYALE}
\mbox{M.Morii\unskip,\iSLAC}
\mbox{D. Muller\unskip,\iSLAC}
\mbox{V. Murzin\unskip,\iMOSCOW}
\mbox{T. Nagamine\unskip,\iTOHO}
\mbox{S. Narita\unskip,\iTOHO}
\mbox{U. Nauenberg\unskip,\iCOLO}
\mbox{H. Neal\unskip,\iSLAC}
\mbox{M. Nussbaum\unskip,\iCINC}
\mbox{N. Oishi\unskip,\iNAGO}
\mbox{D. Onoprienko\unskip,\iTENN}
\mbox{L.S. Osborne\unskip,\iMIT}
\mbox{R.S. Panvini\unskip,\iVAND}
\mbox{C.H. Park\unskip,\iSOONG}
\mbox{T.J. Pavel\unskip,\iSLAC}
\mbox{I. Peruzzi\unskip,\iFRAS}
\mbox{M. Piccolo\unskip,\iFRAS}
\mbox{L. Piemontese\unskip,\iFERR}
\mbox{K.T. Pitts\unskip,\iOREG}
\mbox{R.J. Plano\unskip,\iRUTG}
\mbox{R. Prepost\unskip,\iWISC}
\mbox{C.Y. Prescott\unskip,\iSLAC}
\mbox{G.D. Punkar\unskip,\iSLAC}
\mbox{J. Quigley\unskip,\iMIT}
\mbox{B.N. Ratcliff\unskip,\iSLAC}
\mbox{T.W. Reeves\unskip,\iVAND}
\mbox{J. Reidy\unskip,\iMISSI}
\mbox{P.L. Reinertsen\unskip,\iUCSC}
\mbox{P.E. Rensing\unskip,\iSLAC}
\mbox{L.S. Rochester\unskip,\iSLAC}
\mbox{P.C. Rowson\unskip,\iCOLU}
\mbox{J.J. Russell\unskip,\iSLAC}
\mbox{O.H. Saxton\unskip,\iSLAC}
\mbox{T. Schalk\unskip,\iUCSC}
\mbox{R.H. Schindler\unskip,\iSLAC}
\mbox{B.A. Schumm\unskip,\iUCSC}
\mbox{J. Schwiening\unskip,\iSLAC}
\mbox{S. Sen\unskip,\iYALE}
\mbox{V.V. Serbo\unskip,\iSLAC}
\mbox{M.H. Shaevitz\unskip,\iCOLU}
\mbox{J.T. Shank\unskip,\iBU}
\mbox{G. Shapiro\unskip,\iLBL}
\mbox{D.J. Sherden\unskip,\iSLAC}
\mbox{K.D. Shmakov\unskip,\iTENN}
\mbox{C. Simopoulos\unskip,\iSLAC}
\mbox{N.B. Sinev\unskip,\iOREG}
\mbox{S.R. Smith\unskip,\iSLAC}
\mbox{M.B. Smy\unskip,\iCSU}
\mbox{J.A. Snyder\unskip,\iYALE}
\mbox{H. Staengle\unskip,\iCSU}
\mbox{A. Stahl\unskip,\iSLAC}
\mbox{P. Stamer\unskip,\iRUTG}
\mbox{H. Steiner\unskip,\iLBL}
\mbox{R. Steiner\unskip,\iADEL}
\mbox{M.G. Strauss\unskip,\iMASS}
\mbox{D. Su\unskip,\iSLAC}
\mbox{F. Suekane\unskip,\iTOHO}
\mbox{A. Sugiyama\unskip,\iNAGO}
\mbox{S. Suzuki\unskip,\iNAGO}
\mbox{M. Swartz\unskip,\iJHU}
\mbox{A. Szumilo\unskip,\iWASH}
\mbox{T. Takahashi\unskip,\iSLAC}
\mbox{F.E. Taylor\unskip,\iMIT}
\mbox{J. Thom\unskip,\iSLAC}
\mbox{E. Torrence\unskip,\iMIT}
\mbox{N.K. Toumbas\unskip,\iSLAC}
\mbox{T. Usher\unskip,\iSLAC}
\mbox{C. Vannini\unskip,\iPISA}
\mbox{J. Va'vra\unskip,\iSLAC}
\mbox{E. Vella\unskip,\iSLAC}
\mbox{J.P. Venuti\unskip,\iVAND}
\mbox{R. Verdier\unskip,\iMIT}
\mbox{P.G. Verdini\unskip,\iPISA}
\mbox{D.L. Wagner\unskip,\iCOLO}
\mbox{S.R. Wagner\unskip,\iSLAC}
\mbox{A.P. Waite\unskip,\iSLAC}
\mbox{S. Walston\unskip,\iOREG}
\mbox{S.J. Watts\unskip,\iBRUN}
\mbox{A.W. Weidemann\unskip,\iTENN}
\mbox{E. R. Weiss\unskip,\iWASH}
\mbox{J.S. Whitaker\unskip,\iBU}
\mbox{S.L. White\unskip,\iTENN}
\mbox{F.J. Wickens\unskip,\iRAL}
\mbox{B. Williams\unskip,\iCOLO}
\mbox{D.C. Williams\unskip,\iMIT}
\mbox{S.H. Williams\unskip,\iSLAC}
\mbox{S. Willocq\unskip,\iMASS}
\mbox{R.J. Wilson\unskip,\iCSU}
\mbox{W.J. Wisniewski\unskip,\iSLAC}
\mbox{J. L. Wittlin\unskip,\iMASS}
\mbox{M. Woods\unskip,\iSLAC}
\mbox{G.B. Word\unskip,\iVAND}
\mbox{T.R. Wright\unskip,\iWISC}
\mbox{J. Wyss\unskip,\iPADO}
\mbox{R.K. Yamamoto\unskip,\iMIT}
\mbox{J.M. Yamartino\unskip,\iMIT}
\mbox{X. Yang\unskip,\iOREG}
\mbox{J. Yashima\unskip,\iTOHO}
\mbox{S.J. Yellin\unskip,\iUCSB}
\mbox{C.C. Young\unskip,\iSLAC}
\mbox{H. Yuta\unskip,\iAOMORI}
\mbox{G. Zapalac\unskip,\iWISC}
\mbox{R.W. Zdarko\unskip,\iSLAC}
\mbox{J. Zhou\unskip.\iOREG}

\it
  \vskip \baselineskip                   
  \centerline{(The SLD Collaboration)}   
  \vskip \baselineskip        
  \baselineskip=.75\baselineskip   
\iADEL
  Adelphi University, Garden City, New York 11530, \break
\iAOMORI
  Aomori University, Aomori , 030 Japan, \break
\iBOLO
  INFN Sezione di Bologna, I-40126, Bologna, Italy, \break
\iBRI
  University of Bristol, Bristol, U.K., \break
\iBRUN
  Brunel University, Uxbridge, Middlesex, UB8 3PH United Kingdom, \break
\iBU
  Boston University, Boston, Massachusetts 02215, \break
\iCINC
  University of Cincinnati, Cincinnati, Ohio 45221, \break
\iCOLO
  University of Colorado, Boulder, Colorado 80309, \break
\iCOLU
  Columbia University, New York, New York 10533, \break
\iCSU
  Colorado State University, Ft. Collins, Colorado 80523, \break
\iFERR
  INFN Sezione di Ferrara and Universita di Ferrara, I-44100 Ferrara, Italy, \break
\iFRAS
  INFN Lab. Nazionali di Frascati, I-00044 Frascati, Italy, \break
\iILLI
  University of Illinois, Urbana, Illinois 61801, \break
\iJHU
  Johns Hopkins University,  Baltimore, Maryland 21218-2686, \break
\iLBL
  Lawrence Berkeley Laboratory, University of California, Berkeley, California 94720, \break
\iLTU
  Louisiana Technical University, Ruston,Louisiana 71272, \break
\iMASS
  University of Massachusetts, Amherst, Massachusetts 01003, \break
\iMISSI
  University of Mississippi, University, Mississippi 38677, \break
\iMIT
  Massachusetts Institute of Technology, Cambridge, Massachusetts 02139, \break
\iMOSCOW
  Institute of Nuclear Physics, Moscow State University, 119899, Moscow Russia, \break
\iNAGO
  Nagoya University, Chikusa-ku, Nagoya, 464 Japan, \break
\iOREG
  University of Oregon, Eugene, Oregon 97403, \break
\iOXF
  Oxford University, Oxford, OX1 3RH, United Kingdom, \break
\iPADO
  INFN Sezione di Padova and Universita di Padova I-35100, Padova, Italy, \break
\iPERU
  INFN Sezione di Perugia and Universita di Perugia, I-06100 Perugia, Italy, \break
\iPISA
  INFN Sezione di Pisa and Universita di Pisa, I-56010 Pisa, Italy, \break
\iRAL
  Rutherford Appleton Laboratory, Chilton, Didcot, Oxon OX11 0QX United Kingdom, \break
\iRUTG
  Rutgers University, Piscataway, New Jersey 08855, \break
\iSLAC
  Stanford Linear Accelerator Center, Stanford University, Stanford, California 94309, \break
\iSOGA
  Sogang University, Seoul, Korea, \break
\iSOONG
  Soongsil University, Seoul, Korea 156-743, \break
\iTENN
  University of Tennessee, Knoxville, Tennessee 37996, \break
\iTOHO
  Tohoku University, Sendai 980, Japan, \break
\iUCSB
  University of California at Santa Barbara, Santa Barbara, California 93106, \break
\iUCSC
  University of California at Santa Cruz, Santa Cruz, California 95064, \break
\iUVIC
  University of Victoria, Victoria, British Columbia, Canada V8W 3P6, \break
\iVAND
  Vanderbilt University, Nashville,Tennessee 37235, \break
\iWASH
  University of Washington, Seattle, Washington 98105, \break
\iWISC
  University of Wisconsin, Madison,Wisconsin 53706, \break
\iYALE
  Yale University, New Haven, Connecticut 06511. \break

\rm
%

\end{center}

\end{document}